\documentclass{article}

\usepackage[final]{generic_conf}

\usepackage{amsmath,amsfonts,bm}

\def\eqref#1{equation~\ref{#1}}
\def\1{\bm{1}}

\DeclareMathAlphabet{\mathsfit}{\encodingdefault}{\sfdefault}{m}{sl}
\SetMathAlphabet{\mathsfit}{bold}{\encodingdefault}{\sfdefault}{bx}{n}

\usepackage[utf8]{inputenc} % allow utf-8 input
\usepackage[T1]{fontenc}    % use 8-bit T1 fonts
\usepackage{booktabs}       % professional-quality tables
\usepackage{amsfonts}       % blackboard math symbols
\usepackage{nicefrac}       % compact symbols for 1/2, etc.
\usepackage{microtype}      % microtypography
\usepackage{amsmath}
\usepackage{amssymb}
\usepackage{bbm}
\usepackage{graphicx}
\usepackage{xcolor}
\usepackage{url}
\usepackage{hyperref}
\usepackage{natbib}

\newcommand\sinead[1]{{\color{red}{\bf Sinead: #1}}}

\newcommand\linearItemEmbedding[1]{\mathbf{w}_{#1}}
\newcommand\nonlinearItemEmbedding[1]{\mathbf{x}_{#1}}
\newcommand\nonlinearItemContext[1]{\mathbf{y}_{#1}}
\newcommand\linearUserEmbedding[1]{\mathbf{v}_{#1}}
\newcommand\nonlinearUserEmbedding[1]{f(#1)}
\newcommand\genericWeight{\phi}

\newcommand\UU{\mathcal{U}}
\newcommand\A{\mathcal{A}}
\newcommand\futureSet{\mathcal{S}}
\newcommand\observedSet{\mathcal{T}}
\newcommand\negativeSet{\mathcal{N}}
\newcommand\positiveSet{\mathcal{S}}

\title{Large-scale Collaborative Filtering with Product Embeddings}

\author{
  Thom Lake \\
  Amazon.com \\
  \texttt{thomlake@amazon.com} \\
  \And
  Sinead A. Williamson \\
  Amazon.com, \\
  IROM/Statistics and Data Science \\
  University of Texas at Austin \\
  \texttt{sinead.williamson@mccombs.utexas.edu} \\
  \And
  Alexander T. Hawk \\
  Amazon.com \\
  \texttt{ahhawk@amazon.com} \\
  \And
  Christopher C. Johnson \\
  Amazon.com \\
  \texttt{nchrij@amazon.com} \\
  \And
  Benjamin P. Wing \\
  Amazon.com \\
  \texttt{benjwing@amazon.com} \\
}

\begin{document}

\maketitle

\begin{abstract}
The application of machine learning techniques to large-scale personalized recommendation problems is a challenging task. Such systems must make sense of enormous amounts of implicit feedback in order to understand user preferences across numerous product categories. This paper presents a deep learning based solution to this problem within the collaborative filtering with implicit feedback framework. Our approach combines neural attention mechanisms, which allow for context dependent weighting of past behavioral signals, with representation learning techniques to produce models which obtain extremely high coverage, can easily incorporate new information as it becomes available, and are computationally efficient. Offline experiments demonstrate significant performance improvements when compared to several alternative methods from the literature. Results from an online setting show that the approach compares favorably with current production techniques used to produce personalized product recommendations.
\end{abstract}

\section{INTRODUCTION}\label{sec:intro}

Helping users discover content and products in increasingly complex digital marketplaces is crucial for e-commerce companies like Amazon. In order to accomplish this task, large-scale recommendation systems must model the preference relationships between millions of users and items. Collaborative filtering methods, which aim to model user preferences based on previous user-item interactions, are frequently employed to solve such problems \citep{amazon-item-item}. In particular, latent factor models, such as those based on matrix factorization, have seen widespread success \citep{does-content-matter}.

Unfortunately, traditional recommender systems that learn embeddings for both users and items \citep{mf-survey, imf} scale at best linearly with the combined number of users and items, and are not appropriate when the numbers of users and items can be in the tens or hundreds of millions (in this work, 40 million users and 10 million items). In such large-scale settings, matrix factorization based approaches to collaborative filtering require sophisticated techniques to overcome the computational and memory strains they pose \citep{large-scale-mf, gpu-mf}.

Furthermore, in real-world settings, numerous related problems which require understanding items, users, and preferences are likely being worked on simultaneously. Repeatedly solving such problems from scratch introduces operational inefficiencies and slows rates of innovation due to the diversion of resources from dealing with the unique aspects of an individual domain. While it may be possible to reclaim these resources by limiting scope, for example, building category specific recommender systems that simply ignore user-item interactions deemed to be out of domain, doing so introduces new to category cold-start problems \citep{cold-start} that would not exist if one had modeled a broader view of user tastes and preferences.

% One way to overcome the deficiencies outlined above is to decompose the recommendation problem into distinct item representation and user preference modelling phases. To learn representations of items, any suitable representation learning algorithm could be used \citep{prod2vec,mrnet,starspace,meta-prod2vec}. Here, we employ well-known algorithms, originally developed for learning distributed representations of words \citep{word2vec-mikolov}, which have recently been adapted to the collaborative filtering setting \citep{airbnb}. These pre-trained item representations can be reused to quickly bootstrap machine learning approaches for a variety of e-commerce related tasks such as similarity search, deduplication, or user preference modeling. The generic vector based form of the representations allows them to be easily shared by researchers and practitioners and provide for a clear point of delineation between systems. As a result, any improvements made to the representation learning framework can immediately benefit downstream tasks without requiring the refactoring of complex production machine learning pipelines \citep{from-the-trenches,nuts-and-bolts}. However, to the best of our knowledge, methods of using pre-trained item embeddings for user preference modeling have not been extensively studied.

One way to overcome the deficiencies outlined above is to decompose the recommendation problem into distinct item representation and user preference modeling phases. Reusing a shared set of highly informative pre-trained item embeddings significantly reduces computational costs, and allows us to focus our efforts on experimentally measuring the effectiveness of downstream collaborative filtering models. While there has been a number of item embedding algorithms proposed in the literature \citep{airbnb,prod2vec,mrnet,starspace,meta-prod2vec}, to the best of our knowledge, methods of leveraging such pre-trained item embeddings for user preference modeling have not been extensively studied.

In this work, we explore a number of methods which use pre-trained item embeddings within the large-scale collaborative filtering within implicit feedback setting. Our primary contribution is the novel use of an attention mechanism to select relevant pieces of information from a user’s historical behavior given a particular query item. As such, the proposed method is better described as jointly representing the compatibility between a specific user-item pair, rather that learning distinct user and item representations in a common latent space. We demonstrate that this architecture is extremely well-suited to leveraging such pre-trained distributed representations of items for user preference modeling.

To learn representations of items, any suitable representation learning algorithm could be used \citep{prod2vec,mrnet,starspace,meta-prod2vec}. In this work, we employ well-known algorithms, originally developed for learning distributed representations of words \citep{word2vec-mikolov}, which have recently been adapted to the collaborative filtering setting \citep{airbnb}. These pre-trained item representations can be reused to quickly bootstrap machine learning approaches for a variety of e-commerce related tasks such as similarity search, deduplication, or user preference modeling. The generic vector based form of the representations allows them to be easily shared by researchers and practitioners, and provides for a clear point of delineation between systems. As a result, any improvements made to the representation learning framework can immediately benefit downstream tasks without requiring the refactoring of complex production machine learning pipelines \citep{from-the-trenches,nuts-and-bolts}.

After providing background and reviewing related work in Section~\ref{sec:background}, we describe the components of our approach in Sections~\ref{sec:representations} and \ref{sec:abcf}. We demonstrate their efficacy on large-scale offline experiments using Amazon user-item interaction data in Section~\ref{sec:experiments}, and a production setting in Section~\ref{sec:tp-cards}. In both cases, we show impressive performance against a range of alternative approaches.

\section{BACKGROUND AND RELATED WORK}\label{sec:background}

% Our model builds upon a rich body of work in collaborative filtering and representation learning. Collaborative filtering systems for implicit feedback data attempt to predict future user behavior based on previously observed interactions. Let $\UU$ and $\A$ be sets of users and items respectively. For user $u \in \UU$ and item $i \in \A$, let $e_{ui} = 1$ if user $u$ has interacted with item $i$, and $e_{ui} = 0$ otherwise. The set of all observed item interactions for user $u$ is denoted $\A_u = \{i \in \A \colon e_{ui} = 1\}$. The collaborative filtering task is then to predict $e_{uj}$ for previously unseen items $j$.

Our model builds upon a rich body of work in collaborative filtering and representation learning. Collaborative filtering systems for implicit feedback data attempt to predict future user behavior based on previously observed interactions. Let $\UU$ and $\A$ be sets of users and items respectively. For user $u \in \UU$ and item $i \in \A$, let $e_{ui} = 1$ if user $u$ has interacted with item $i$, and $e_{ui} = 0$ otherwise. The set of all observed item interactions for user $u$ is denoted $\A_u = \{i \in \A \colon e_{ui} = 1\}$. The collaborative filtering task is then to predict $e_{uj}$ for previously unseen items $j$.

\subsection{Latent Factor based Approaches}

A commonly employed approach to collaborative filtering is to associate vectors of latent factors $\linearUserEmbedding{u}, \linearItemEmbedding{i} \in \mathbb{R}^d$ with each user and item. The degree to which some query item $q \in \mathcal{A}$ is relevant for a specific user is taken to be a function of the inner product between these vectors \citep{mf-survey},

\begin{align} \label{eq:mf-score}
  r(u, q) &= \linearUserEmbedding{u}^T \linearItemEmbedding{q}.
\end{align}

We can then formulate the problem as recovering the observed interactions and estimate parameters by minimizing some suitable loss function,

\begin{align*}
  % \label{eq:cf-loss}
  \mathcal{L}(\theta) = \sum_{u\in \UU} \sum_{i \in \A}\ell \left(e_{ui}, r(u, i)\right).
\end{align*}

Common choices of loss include weighted least squares \citep{imf} or the logistic loss \citep{logistic-imf}.

While such linear embedding approaches to collaborative filtering perform well in a variety of settings, there are a number of factors that make it difficult to apply in a large-scale recommendation setting like the one considered here.

First and foremost is the sheer number of parameters one must estimate. For example, the dataset used in this work consists of 40 million users and 10 million items. Using $d = 64$ dimensional embeddings would require estimating over three billion parameters.

 % for example, 64 dimensional embeddings would require over 3B parameters (≈2.5B for users, ≈500M for items)

Furthermore, since the number of observed interactions tends to grow linearly with the number of users, the user-item interaction matrix becomes increasingly sparse. Latent factor models operating in this regime require careful regularization to ensure models generalize appropriately to unseen interactions \citep{prob-matrix-fact}.

Lastly, since latent factor models are transductive in nature, i.e.\ they do not generalize to new users or items, rapidly incorporating new information into the system can be problematic. For example, incorporating new item interactions into the user representation typically requires solving an optimization problem \citep{logistic-imf}.

\subsection{Learning Nonlinear Encodings of users}

The transductive nature of latent factor models can be ameliorated by replacing the linear embedding, $\linearUserEmbedding{u}$, in (\ref{eq:mf-score}) with a nonlinear embedding of user features,

\begin{align}
  \label{eq:nn-cf}
  r(u, q) &= \nonlinearUserEmbedding{u}^T \linearItemEmbedding{q}.
\end{align}

A number of deep learning based recommendation systems have been proposed, that use a neural network to capture this nonlinear embedding \citep{neural-matrix-factorization,rbm-cf}. A common focus of these works is devising methods which are able to exploit various types of side information which are difficult to incorporate when using tranditional matrix factorization based techniques. In recommendation contexts, such side information often requires special handling as it may be very sparse (search queries, geographic location) \citep{wide-and-deep,deep-and-cross,deepfm}, or temporally structured (session or browsing history) \citep{session-based-rnn}. 

% Due to the parametric nature of the user representations, neural network based recommender systems can easily incorporate new information about a user, and may exhibit superior generalization capabilities.

% A number of deep learning based recommendation systems have been proposed, that use a neural network to capture this nonlinear embedding \citep{neural-matrix-factorization,rbm-cf}. A common focus of these works is devising methods which are able to exploit various types of side information which are difficult to incorporate within the matrix factorization framework. In recommendation contexts, such side information often requires special handling as it may be very sparse (search queries, geographic location) \citep{wide-and-deep,deep-and-cross,deepfm}, or temporally structured (session or browsing history) \citep{session-based-rnn}. Due to the parametric nature of the user representations, neural network based recommender systems can easily incorporate new information about a user online, and may exhibit superior generalization capabilities.

Historical user behavior is commonly incorporated into deep learning based recommendation systems using a \emph{bag-of-words} style encoding of past user-item interactions, i.e, summing or averaging item embeddings \citep{autorec,youtube-google,neural-autoregressive}. Several recent works have explicitly explored order-independent combination operators operating over sets \citep{deep-avg-net,deep-sets,relational-networks}, and surprisingly good performance can be achieved using simple sums, averages, and TF-IDF style weighted averages of word embeddings \citep{sentence-embedding-meaning,weighted-word-embeddings} in an NLP context. The use of vector averaging is especially appealing due its simplicity, ability to deal with variable sized user behavior histories, and the ease with which new information can be incorporated.

However, as we show experimentally in Section~\ref{sec:experiments}, using a static combination operation may be suboptimal in some settings and significant improvements can be achieved using learned combination operators that can adapt to the context at hand.

\subsection{Nonlinear Functions of Items and users}

Despite modeling users using flexible nonlinear embeddings, most deep learning based recommender systems share the common architectural feature of modeling relevance as the inner product between a separate user and item vector, as in (\ref{eq:nn-cf}). While this may be sufficient given a large enough number of factors and training dataset, simply enlarging this latent space may be suboptimal both from a generalization and a computational perspective. An alternative, which is explored in this work, is to assume that relevance is some nonlinear function of both the user and the query item.

This idea has been explored by a number of authors. \cite{multiple-users} propose learning multiple vectors for each user and select among them using a max operation. A learned weighting of individual user and item factors is proposed in \cite{neural-cf}. In both cases, these models require learning distinct representations of users and suffer from similar problems as matrix factorization based approaches. Closer in spirit to our work is that of \cite{meta-learning-recs}, which proposes a deep learning based recommender system motivated by a meta-learning perspective \citep{lake-meta-learning}. Their model outputs a relevance score given a query item vector and two user vectors computed by averaging the vectors for items in the sets $\A_u$ and $\A \setminus \mathcal{A}_u$. Our approach offers greater flexibility, moving beyond simple averages and using a more flexible functional form.

Another related approach was recently explored by \citet{airbnb}. Like the work presented here, the techniques presented in \citet{airbnb} rely on item representations learned using a \texttt{word2vec} \citep{word2vec-mikolov} like procedure. Unlike this work, \citeauthor{airbnb}'s training procedure simultaneously learns a \emph{user type} representation while learning item embeddings. Specific users are then mapped to user types by manually derived rules. Ultimately, these representations are used to derive relevance features (cosine distances) which are input into a decision tree based personalization model. In contrast, our approach leverages item embeddings directly to model user prefeences by learning a parametrized model operating over the items a user has previously interacted with. Decoupling these procedures and treating the embeddings as inputs sidesteps the need to define heuristic rules to map from users to types, and allows models to easily incorporate new behavioral signals without retraining.

\subsection{Neural Attention Mechanisms for Combining Representations}

In addition to summing or averaging item representations, we explore the use of neural attention mechanism to combine item representations. Neural attention mechanisms \citep{olah-attention, ntm, attention-align-and-translate, structured-attention} provide a method by which neural networks can focus processing on particular portions of the input data and ignore those which may be irrelevant in the current context. Although most commonly used to augment recurrent neural networks, attention mechanisms can also be used as an alternative to convolution \citep{cnn-sentence} or recurrence when dealing with variable size structured inputs such as sequences, trees, or sets \citep{attention-is-all-you-need}. The model we propose is related to this line of work, largely inspired by Memory-Networks and their derivatives \citep{memory-networks, end-to-end-memory-networks}.

The use of attention mechanisms specifically in recommender system contexts has also been recently explored. In \citep{attentive-cf}, an attention mechanism is used to select the most important historical item interactions given a particular user. Because the attention mechanism conditions explicitly on the user rather than the query item, the attention weights applied to past item interactions, and the resulting user representation, remain constant for a particular user. More similar to our work, is the concurrent approach of \citep{attention-based-cf}, which applies a matching network \citep{matching-networks} like averaging of past item ratings to predict ratings for unseen items. Unlike our work, \citep{attention-based-cf} introduce an additional matrix factorization component to improve performance. Thus, both methods require learning an embedding for each user, something we explictly avoid for the reasons described in Section \ref{sec:intro}.

% The model they propose applies attention weights, conditioned on a context item, on past item interactions. However, \citep{attention-based-cf} apply their model rating prediction, and use the attention weights to produce an averaged rating, using a method similar to \cite{matching-networks},

% This requires the model to learn a unique embedding for each user, something we explicitly avoid for reasons described previously. Furthermore, in ,   As such, there are effectively $|\A|$ representations per user (one for each item). For this reason, the model we propose is better described as jointly representing the compatability between a specific user-item pair, rather that learning distinct user and item representations in a common latent space.

% \sinead{I'm leaving Alex's comment below because I think it should be addressed, but I think I would put it in section 5 rather than here.}
% \alex{I feel we need to do more to motivate the choice of using an order independent algorithm. Given that we have the order of a users actions, and that we know the order to be highly predictive of what a user will do next, why would we ignore it? Are we ignoring it for the sake of a) simplifying the prediction algorithm or training procedure? b) understanding how good an order independent algorithm is? c) making training more efficient / robust?}

% \section{PRE-TRAINED, GENERAL PURPOSE PRODUCT REPRESENTATIONS}\label{sec:representations}
\section{PRODUCT REPRESENTATIONS}\label{sec:representations}

Our nonlinear collaborative filtering method, described in Section~\ref{sec:abcf}, takes as input a set of pre-trained item embeddings. In the context of this paper, items correspond to products available on an e-commerce site. Numerous techniques for learning product embeddings have been explored in the literature. A common theme is for these methods to focus on representing products via a specific modality such as images \citep{joint-representation-products}, text \citep{mrnet}, audio \citep{spotify-cnn}, or combinations thereof. In our setting, we do not assume access to side-information, so representations are learned based purely on user/item interaction data. If such features were also available, this information can easily be incorporated into the item representation learning process \citep{meta-prod2vec, meta-learning-recs}.

In particular, the embeddings used in this work were created using a straightforward adaptation of the skip-gram method of \cite{word2vec-mikolov} to the product embedding setting. This is done by replacing sequences of words with temporally ordered sequences of user actions \citep{meta-prod2vec}. Each sequence consists of a single user's purchases, product views, and digital video and music streams. A window slides over each sequence and co-occurring pairs within the window are recorded. For each item $i$, this yields a neighborhood $\positiveSet_i$ of co-occurring context items. In addition to the positive examples of co-occurring context items in $\positiveSet_i$, we sample a collection $\negativeSet_i$ of \emph{negative samples} randomly from some distribution over items. We wish to learn embeddings that map frequently co-occurring items to similar locations in the embedding space, while maintaining separation between dissimilar items. This is achieved by optimizing the parameters of the embedding to minimize
\begin{align*}
  % \label{eq:skip-gram-loss}
  % \begin{multline}
  \mathcal{L}(\theta) = -\sum_{i} & \left[ \; \sum_{j\in \positiveSet_i} \log \sigma(\nonlinearItemEmbedding{i}^T\nonlinearItemContext{j})
  + \sum_{j^\prime\in \negativeSet_i} \log(1 - \sigma(\nonlinearItemEmbedding{i}^T \nonlinearItemContext{j^\prime})) \right],
  % \end{multline}
\end{align*}
where $\nonlinearItemEmbedding{i}, \nonlinearItemContext{j} \in \mathbb{R}^d$ are the embeddings for item $i$ and context $j$, and $\sigma(\cdot)$ is the logistic function. The resulting embeddings are used as inputs to all the models presented in the remainder of this work. Importantly, the embeddings are left fixed and not updated when training downstream collaborative filtering models. 

An alternative to using pre-trained would be to jointly learn these embeddings as part of our recommender model \citep{neural-autoregressive, autorec, youtube-google}. However, such an approach would significantly increase the complexity of the inference problem, making it infeasible for our large-scale setting. For example, the experiments in Section~\ref{sec:experiments} use a relatively compact model containing just under 200,000 trainable parameters. In comparison, the number of parameters in the item embedding matrix is over 500 million. Splitting the two components simplifies the problem space, and allows the two model components to be learned independently. This is particularly important in large-scale recommendation settings where models must be continuously updated.

While this choice does limit model capacity, the degree to which this will actually hinder performance is unclear. For example, previous results in NLP demonstrated significant performance improvements when using static word embeddings as input, rather than randomly initialized words embeddings trained jointly with the model \citep{cnn-sentence}. The use of fixed embeddings also helps to ensure differences in experimental measurements are due to differences in architecture, and not the result of effects such as over specialization of the embeddings during training (a problem has the the potential to be particularly acute given the extremely large number of items in our dataset). For these reasons, this paper focuses exclusively on the use of static item representations and leaves further exploration of this issue as future work.

\section{COLLABORATIVE FILTERING WITH ATTENTION}\label{sec:abcf}

Many large-scale e-commerce datasets feature user interactions with a variety of item types and categories. Tasked with determining if a given item is relevant to a user in this setting, it would be reasonable to assume that only a small subset of the user's past interactions provide useful information about the problem. Further, the appropriate subset is likely to vary across items. A representation of past item interactions produced using a static combination operator, such as summation or averaging, does not possess this sort of selectivity. We posit that an alternate combination operator, that uses a query dependent weighted average (attention), will provide a more appropriate inductive bias.

\subsection{Model}

We take a probabilistic approach to ranking and aim to model the probability that user $u \in \UU$ will interact with item $q \in \A$ based on observed historical user-item interactions $\A_u$. The first step in our model is transforming the embeddings of candidate item $q$ and each previously observed item $i \in \A_u$ via parametrized functions $\mathbf{h}$ and $\mathbf{f}$ respectively. The transformed representation of $q$ is then compared to each item a user has previously interacted via a dot product. The result of these comparisons is normalized using the softmax function to yield a vector of attention weights,

\begin{align*}
  % \label{eq:attend-simple}
  p_i = \text{softmax}\left(\mathbf{f}\left(\nonlinearItemEmbedding{i}\right)^T \mathbf{h}\left(\nonlinearItemEmbedding{q}\right)\right),\; i \in \A_u.
\end{align*}

Next, the embeddings of the items in $\A_u$ are reduced to a single $d$-dimensional vector by transforming them using a third parameterized function $\mathbf{g}$, and taking a convex combination with weights $p_i$,

\begin{align*}
  % \label{eq:combine-simple}
  \mathbf{z} = \sum_{i \in \A_i} p_i \mathbf{g}\left(\nonlinearItemEmbedding{i}\right).
\end{align*}

\begin{figure}
  \centering
  \includegraphics[width=0.5\textwidth]{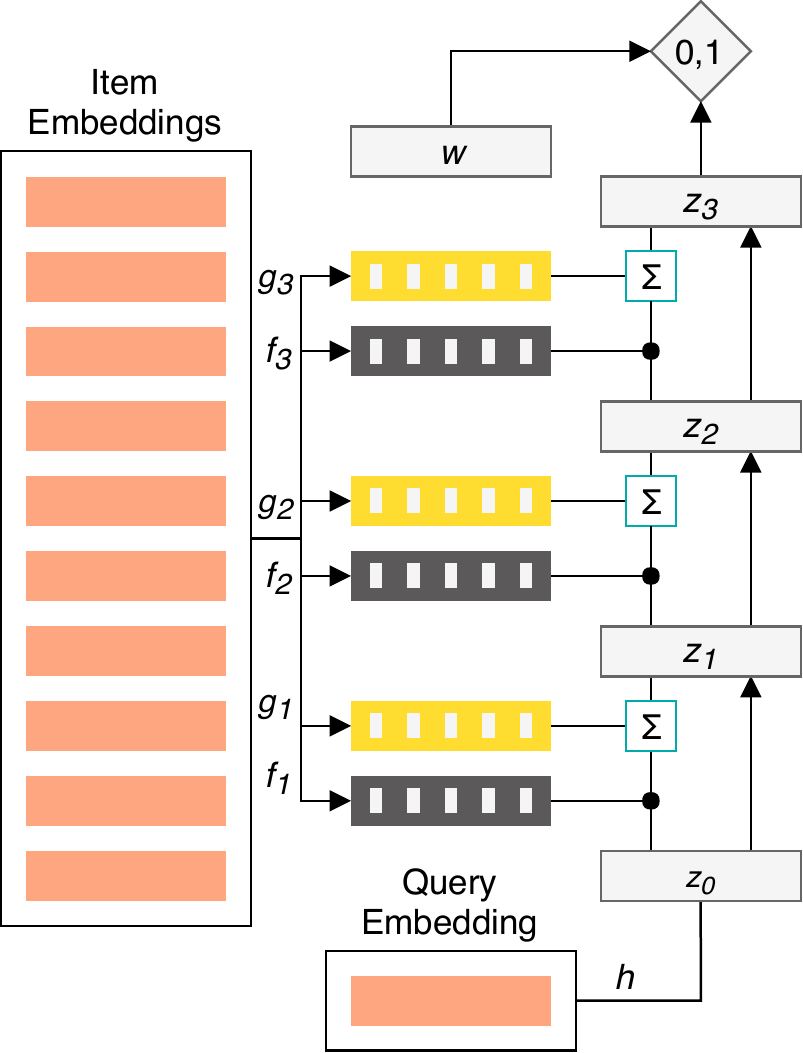}
  \caption{Flow of information through a depth three version of the attention based collaborative filtering model.}
  \label{fig:attend-and-rank}
\end{figure}

Multiple layers of this basic motif are stacked. Information is integrated across layers by adding the query and output of the attention mechanism and using the result as the query in the next layer \citep{end-to-end-memory-networks}. A depth-$K$ version of the resulting architecture may be written as

\begin{align*}
  \mathbf{z}^0 &= \mathbf{h}\left(\nonlinearItemEmbedding{q}\right) \\
  p_i^k &= \text{softmax}\left(\left(\mathbf{f}^k(\nonlinearItemEmbedding{i})\right)^T \mathbf{z}^{k-1}\right) \\
  \mathbf{z}^k &= \mathbf{z}^{k-1} + \sum_{i \in \A_u} p^k_i \left(\mathbf{g}^k(\nonlinearItemEmbedding{i})\right),
\end{align*}

where $k=1,\ldots,K$. The resulting joint representation $\mathbf{z}:=\mathbf{z}^K$ of the query item $q$ and the user's historical item interactions $\A_u$ is reduced to a single score

\begin{align}
  r(q, \A_u) = \sigma(\genericWeight^T \mathbf{z}),\label{eqn:score}
\end{align}

where $\sigma(\cdot)$ is the logistic function, and $\genericWeight \in \mathbb{R}^{d^\prime}$ is a weight vector that is shared across all user/query pairs. Figure \ref{fig:attend-and-rank} provides a visual representation of the described architecture.

The use of distinct $\mathbf{f}$, $\mathbf{g}$, and $\mathbf{h}$ is similar to the Key-Value Memory Network of \cite{key-value-memory-networks}, and allows the underlying item embeddings to be adapted for the various purposes they serve within the model, without requiring us to learn multiple new representations of millions of items. Parameters are not shared between layers, so each may potentially focus on different aspects of similarity. In the experiments presented here, $\mathbf{f}$, $\mathbf{g}$, and $\mathbf{h}$ take the form of single-layer linear neural networks

\begin{align*}
  \mathbf{f}^k, \mathbf{g}^k, \mathbf{h} & \colon \mathbb{R}^d \longrightarrow \mathbb{R}^{d^\prime} \\
  \mathbf{f}^k(\mathbf{x}) &=  \mathbf{B}_{fk} \mathbf{x} + \mathbf{c}_{fk} \\
  \mathbf{g}^k(\mathbf{x}) &=  \mathbf{B}_{gk} \mathbf{x} + \mathbf{c}_{gk} \\
  \mathbf{h}(\mathbf{x}) &=  \mathbf{B}_h \mathbf{x} + \mathbf{c}_h.
\end{align*}

\subsection{Objective}

To optimize the parameters of the model, we temporally partition each user's item interactions $\A_u$ into a set of future and observed item interactions, denoted by $\futureSet_u$ and $\observedSet_u$ respectively. Predicting a set of future interactions from past interactions reduces the problem to multi-label classification, which could be solved by minimizing the negative log-likelihood,
% \begin{equation}
% \begin{aligned}
%   \label{eq:loss}
%   \mathcal{L}(\theta) = {} & -\sum_{u \in \UU} \sum_{q \in \A} \bigg[\mathbbm{1}_{[q \in \futureSet_u]} \log r(q, \observedSet_u)
%   + \mathbbm{1}_{[q \in \A \setminus \futureSet_u]} \log\left(1 - r(q, \observedSet_u)\right) \bigg].
% \end{aligned}
% \end{equation}

\begin{align}
  \label{eqn:loss}
  \mathcal{L}(\theta) =& -\sum_{u \in \UU} \sum_{q \in \A}
   \bigg[\mathbbm{1}\left\{q \in \futureSet_u\right\} \log r(q, \observedSet_u) 
   +\mathbbm{1}\left\{q \in \A \setminus \futureSet_u\right\} \log\left(1 - r(q, \observedSet_u)\right) \bigg].
\end{align}

Unfortunately, in practice $|\A|$ is quite large. This makes the sum over each item  in (Eqn.~\ref{eqn:loss}) computationally difficult. To alleviate this issue we employ a negative sampling technique \citep{sampling-strategies-nncf,one-class-cf,word2vec-mikolov}. For each user, we sample a set of negative items, $\negativeSet_u \subset \A$, $\negativeSet_u \cap \futureSet_u = \emptyset$, from a smoothed empirical distribution $P_\gamma$ obtained by raising the probability of each item occurring in the training data to a fractional power $0 < \gamma < 1$ and renormalizing,

\begin{align}
  \label{eqn:neg-dist}
  P_\gamma(i) &= \frac{
    \left(\sum\limits_u \mathbbm{1}\{i \in \A_u\}\right)^\gamma
  }{
    \sum\limits_{i^\prime}\left(\sum\limits_u \mathbbm{1}\{i^\prime \in \A_u\}\right)^\gamma
  } \, .
\end{align}

To sample negatives while training, we employ the alias method, which allows us to sample from a categorical distribution in $O(1)$ time \citep{alias-method, rbm-words}. We also employ a weighting scheme in our loss which gives equal weight to the positive and negative items associated with each training instance. This keeps the scale of the loss independent of the number of positive and negative items that appear in each instance. We found this made tuning hyperparameters easier. Having done this, we arrive at the following loss

\begin{align*}
  \label{eqn:final-loss}
  \hat{L} =
    -\sum_{u \in \UU} &
    \left[
      \frac{1}{2|\futureSet_u|} \sum_{q \in \futureSet_u} \log r(q, \observedSet_u) 
      +  \frac{1}{2|\negativeSet_u|} \sum_{q \in \negativeSet_u} \log(1 - r(q, \observedSet_u))
    \right].
\end{align*}

\subsection{Training Details} \label{sec:training-details}
The propsed model is fully differentiable and can be trained end-to-end using backpropagation. We use the Adam \citep{adam} optimizer with $\beta_1 = 0.9$, $\beta_2 = 0.999$, minibatches of size 64, and an initial learning rate of 0.002. We  scale the gradient vector to have norm 10 when it exceeds this value \citep{grad-clip}. Model performance is assessed on a small holdout set of users every 5,000 updates, and the learning rate is reduced by a factor of 0.8 when the objective does not improve after 5 consecutive assessments \citep{rnn-empricial}. Training is terminated once this occurs 20 times. Parameters are initialized using the orthogonal initialization scheme of \cite{ortho-init}. We use $\gamma = 0.75$ to smooth the empirical sampling distribution, $|\futureSet_u| = 10$ holdout items, and $|\negativeSet_u| = 100$ negative samples for each training instance. Unless otherwise mentioned, we use a depth $K=10$ version of our model with hidden layers of size $d^\prime = 128$. We tune hyper-parameters on the same set of users used for early stopping, although in many cases found that our initial settings gave reasonable performance and did not explore others. The model is implemented in the PyTorch \citep{pytorch} framework. Training takes around 12 hours on a single GPU.

% The model is implemented in the PyTorch \cite{pytorch} framework. Item embeddings are not updated during training. This results in a relatively compact model consisting of only 174,848 trainable parameters. In comparison, the number of parameters in the item embedding matrix is 540,022,080. Leaving the parameters of the item embeddings fixed greatly reduces our computational requirements. Training the proposed model takes around 12 hours on a single GPU of a p3.16xlarge instance.

\section{EXPERIMENTS} \label{sec:experiments}
To thoroughly evaluate the performance of the proposed attention based approach and our modeling choices, we perform a number of large-scale offline experiments using several model settings. We compare our approach to several alternative methods for leveraging item embeddings that have been proposed in the collaborative filtering literature.

\subsection{Datasets}

Training datasets were constructed by randomly selecting 40 million anonymized Amazon users and aggregating their product purchases and digital content streams over a 120 day period. To assess the performance of models we measured their ability to predict product interactions in the 7 days immediately following the end of the training period. We also removed any items that a user previously interacted with during the training period, choosing to focus exclusively on predicting new item interactions. Because all models use the same set of pre-trained item embeddings, we did not constrain the datasets described here such that all test items appear in the training data. The embeddings are trained on significantly larger datasets, which confers extremely high coverage to downstream algorithms.

\subsection{Baselines}
\label{subsec:baselines}

We compare against several baselines inspired by the existing literature. Since our data does not include side-information, we do not compare against content-based or hybrid content/collaborative filtering methods such as \citep{wide-and-deep,deep-and-cross,deepfm,attentive-cf}. Furthermore, we do not compare to methods that require learning an embedding for each user such as \citep{implicit-feedback,logistic-imf,attention-based-cf,neural-cf,neural-matrix-factorization}, as doing so would be computationally difficult and distract from our primary focus, modeling user preferences with a parametric function of a user's past item interactions.

%Several baselines inspired by the existing literature on embeddings and recommendation were used to predict future customer-item interactions.
 % As discussed in Section~\ref{sec:representations}, we did not update item embeddings during training.

\paragraph{Popularity:}
A non-personalized algorithm that ranks items according to their frequency in the training data.

\paragraph{Last Item:}
An item-based approach which ranks items by cosine similarity (in embedding space) with the last item a user interacted with \citep{airbnb}.

\paragraph{Weighted Sum:}
A weighted average of item embeddings a user has previous interacted with. Weights are set using various heuristics based on factors including time, nature, and type of interaction, and tuned using CMA-ES \citep{cma-es}.

% \paragraph{One-sided Matrix Factorization}
% A logistic matrix factorization \cite{logistic-imf} approach using fixed item embeddings. This reduces the optimization problem to solving an independent binary logistic regression problem for each customer. Combining traditional user-item matrix factorization with item-item cooccurrence based factorization has been previously explored in \cite{factorization-meets-embedding}.

\paragraph{Deep Averaging Network (DAN):}
A deep learning model which uses vector averaging to combine embeddings of items a user has previously interacted with \citep{deep-avg-net,meta-learning-recs,youtube-google}. This input is then passed through a standard feedforward neural network in order to make a relevance determination.

Let $\hat{\mathbf{x}}_u$ be the average of item vectors which user $u$ has interacted with, and $\mu(\cdot)$ be some feedforward neural network. We explored two methods for incorporating a user's past item interactions into the model. The first method adds the average vector to the query vector and passes it through a neural network to produce a relevance score, $r(q, \A_u) = \mu\left(\hat{\mathbf{x}}_u + \nonlinearItemEmbedding{q}\right)$. The second takes the relevance score to be the inner item of the query vector with the average vector after being processed by the neural network, $r(q, \mathcal{A}_u) = \mu\left(\hat{\mathbf{x}}_u\right)^T\nonlinearItemEmbedding{q}$. We present results for the latter as we found it performed better in our setting. We used the same training and hyperparameter tuning procedures as described in Section~\ref{sec:training-details}. The best settings we discovered used two layers of 128 rectified linear units.

This baseline is representative of the most commonly used method of incorporating historical user-item interactions in deep learning recommender systems, and allows us to better understand the importance of the attention mechanism in our approach. In particular, the  architecture is nearly identical to that of the Collaborative Filtering Neural Autoregressive Distribution Estimator \citep{implicit-cf-nade,neural-autoregressive}, User-Autorec \citep{autorec}, and Non-Linear Bias Adaptation \citep{meta-learning-recs} recommender systems, modified to use a pre-trained embedding. Concretely, since multiplying a matrix of pre-trained item embeddings by a sparse-binary vector of observed previous interactions is equivalent to simply summing the item embeddings, the first layer of the DAN model is equivalent (up to a scaling constant) to the first layer of these models. We explore multiple layers of nonlinearities in the DAN, as well as several variants of the exact functional form, suggesting it is largely representative of the best-case performance one would obtain using these methods with fixed item embeddings.

\begin{figure*}
  \centering
  \hspace*{-0.05\textwidth}
  \includegraphics[width=1.1\textwidth]{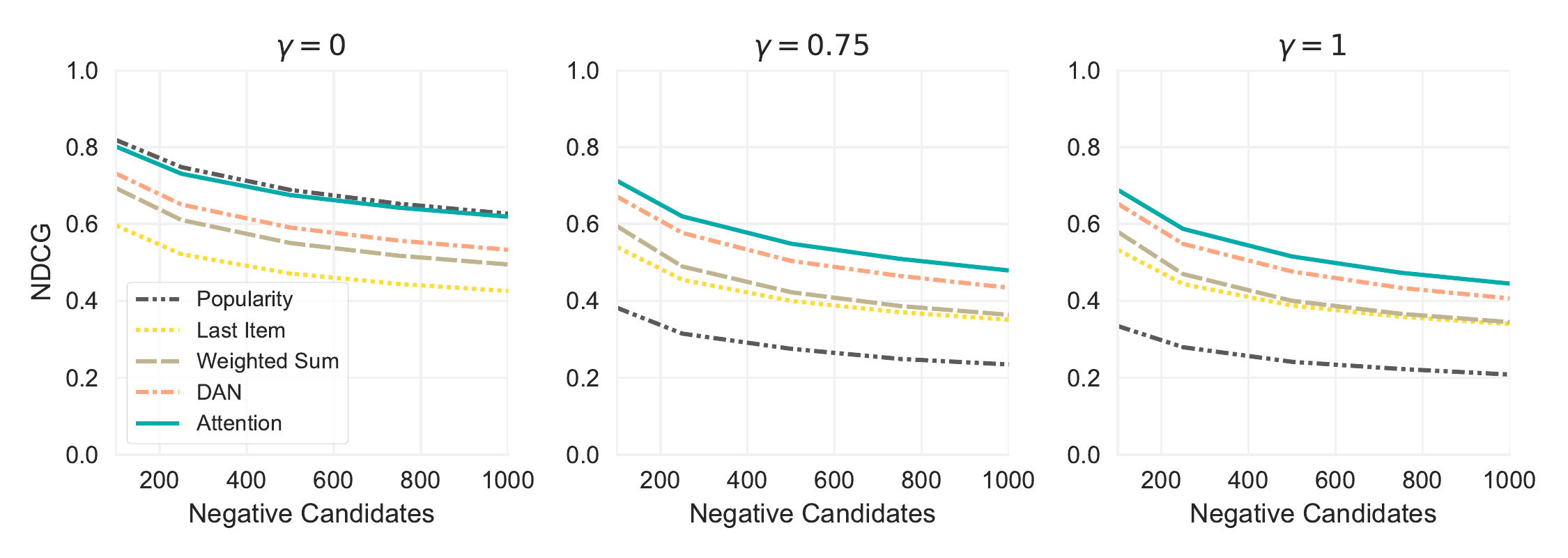}
  \caption{NDCG using various candidates sets. We vary both the number of candidates, and the distribution from which they are drawn.}
  \label{fig:ndcg}
\end{figure*}

\begin{figure*}
  \centering
  \hspace*{-0.05\textwidth}
  \includegraphics[width=1.1\textwidth]{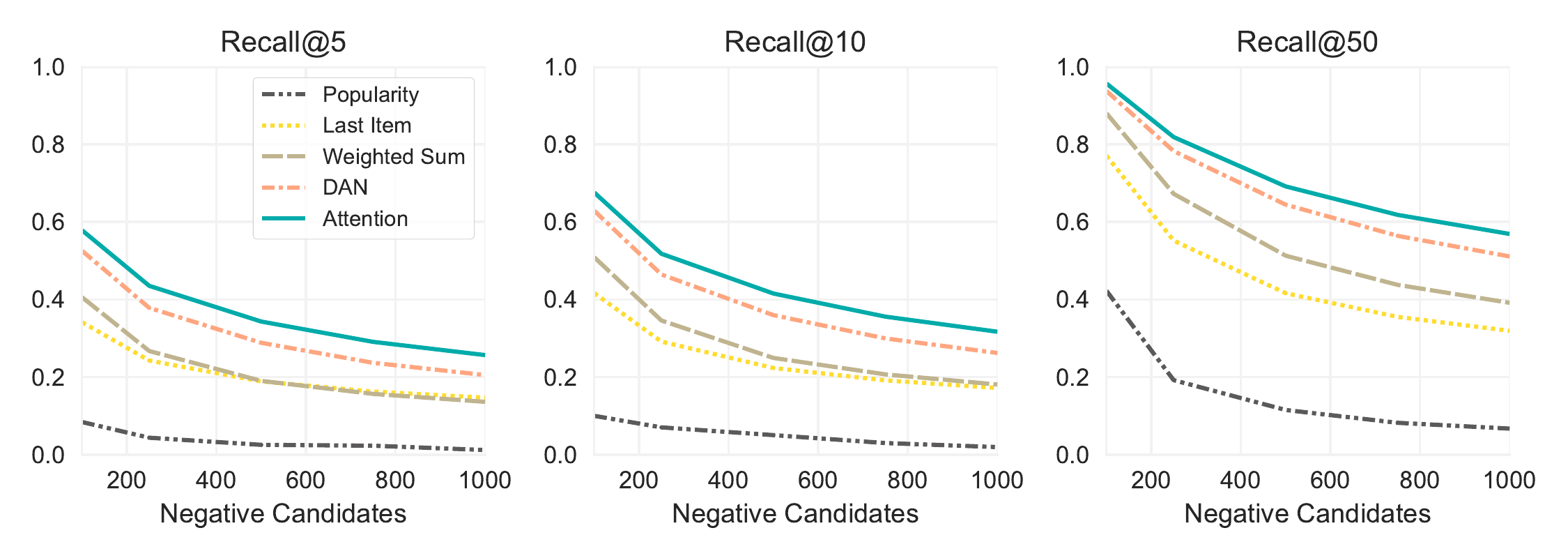}
  \caption{Recall@K of proposed method and baselines ($\gamma=1$).}
  \label{fig:recall}
\end{figure*}

\subsection{Evaluation Metrics}
We evaluated models with normalized discounted cumulative gain (NDCG) \citep{ndcg-theory, ranking-measures} and Recall@K. NDCG allows us to measure the quality of the entire ranked list, while still putting most of the emphasis on items near the top. Recall@K, which has the benefit of being highly interpretable, is the fraction of items ranked amongst the top $K$ that were actually interacted with by the user. 

Because we had to rank an extremely large number of items for each user (our dataset contains several orders of magnitude more items than the widely used Netflix dataset \citep{netflix}), we followed the common practice of ranking the set of items the user interacted with during the test period amongst a set of randomly sampled negative candidate items that were not interacted with \citep{factorization-meets-neighborhood, multi-view-recs, neural-cf}. To explore the implications of this evaluation technique, we vary both the numbers of negative candidates, and the distribution from which the negatives candidates are drawn. For the latter, three different distributions parameterized as in (Eqn.~\ref{eqn:neg-dist}) are used; the empirical distribution ($\gamma=1$), the training distribution ($\gamma=0.75$), and the uniform distribution ($\gamma=0$).

In addition to being computational advantageous, the approach of sampling a pool of candidates is related to the practice of decomposing large-scale recommendation problems into the distinct sub-problems of candidate selection and candidate ranking \citep{youtube-google,rank-and-rerank}. Within this framework, the candidate selection algorithm typically focuses on surfacing generically applicable content with high precision, possibly incorporating various business criteria or eligibility constraints, and the ranking algorithm must order these items for the current user and/or context. From this perspective, all the candidate generation procedures we consider have perfect recall, but differ in terms of approximating an algorithm with high (the empirical distribution) to low (the uniform distribution) precision.

\subsection{Results}
Comparisons against baselines are displayed in Figures \ref{fig:ndcg} and \ref{fig:recall}. In Figure \ref{fig:ndcg} we look at how NDCG varies with both the number of additional negative candidates included in the candidate pool, and the distribution from which they are sampled (recall that more negative candidates implies more candidates to rank, i.e.\ a strictly harder ranking problem). When the sampling distribution for negative candidates resembles that used in training ($\gamma=0.75$) or is set to the empirical distribution ($\gamma=1$), the attention based method outperforms other methods by a clear margin (statistically significant at $\alpha=0.01$), with DAN, our approximation to other state-of-the-art nonlinear recommenders, being the next best performer, and popularity being the worst performer. When negative candidates are sampled from the uniform distribution over items ($\gamma=0$), popularity performs slightly better than the attention based method. However, this performance can be explained by the fact that most items in our dataset are generally ``unpopular'', so sorting by popularity discriminates extremely well between the items that were actually interacted with and a relatively small random sample of items, given the highly skewed nature of user-item interactions.

In Figure~\ref{fig:recall}, we look at Recall@K where the sampling distribution is set to the empirical distribution. Similar results are found when we use $\gamma=0.75$, i.e.\ the training distribution. The findings mirror the results for NDCG: the attention based method achieves the best recall (again, statistically significant at $\alpha=0.01$) and the non-personalized method performs worst. The ability of attention based model to perform well across a variety of settings provides further evidence of its benefits over the alternative techniques considered here.

\begin{figure}[h!]
  \centering
  \includegraphics[width=0.9\textwidth]{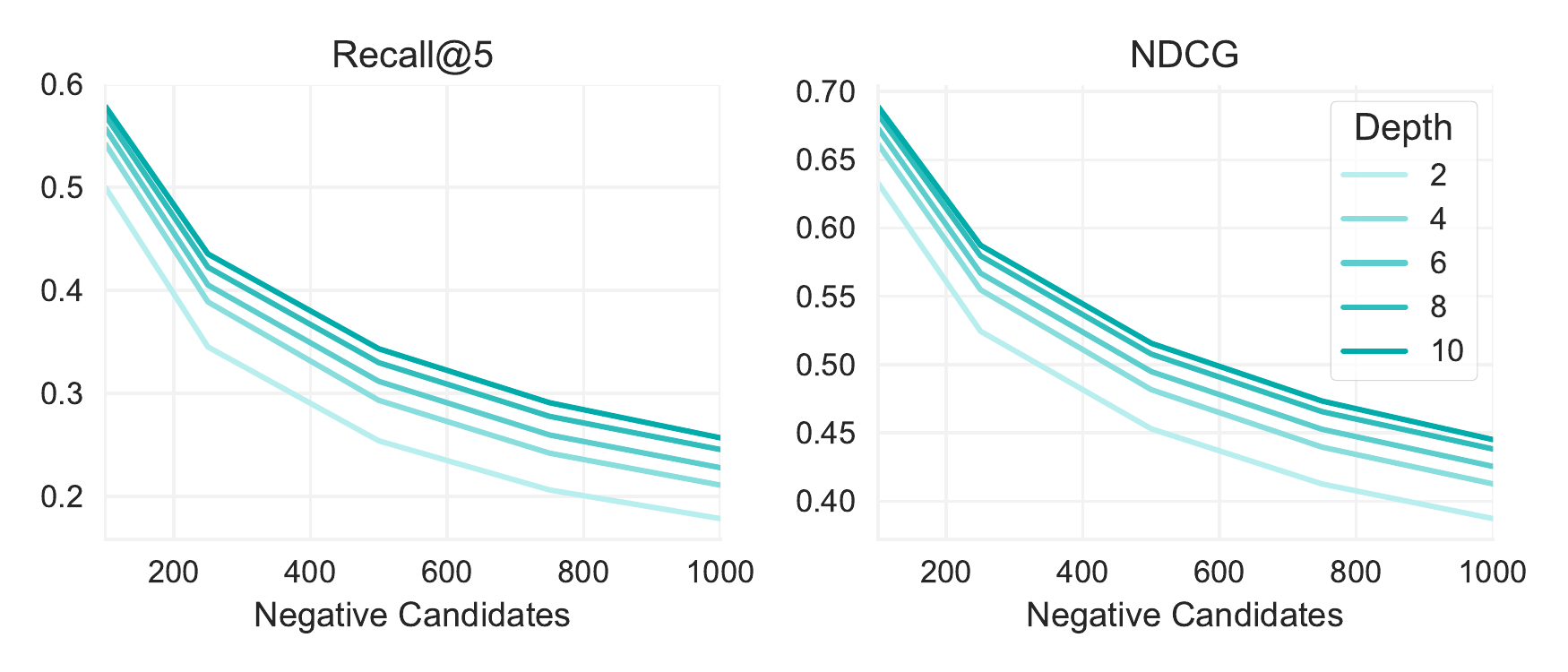}
  \caption{Effect of depth on the attention based model ($\gamma = 1$).}
  \label{fig:depth}
\end{figure}

Figure \ref{fig:depth} shows the effect of increasing depth in the attention based model. Increasing from depth 2 to 4 yields a large improvement in both recall and NDCG. Further increases in depth result in consistent, but diminishing improvements. These results were obtained using negative instances sampled from the empirical distribution. 

\section{CREATING PERSONALIZED CONTENT} \label{sec:tp-cards}

In an e-commerce setting, a key motivation behind inferring customer preferences is to improve customers' experiences. To explore whether the improved preference modeling translates into a high-quality customer experience, we used our proposed attention based model to populate several pieces of content available on the Amazon homepage in a real-world, online setting. Each piece of content featured products selected from one of four categories: Digital Movies, Digital Music, Books, or Video Games. The set of candidate products for each category were chosen in advance according to heuristics that consider attributes such as a product's average rating, listing date, and various business constraints. We used the proposed algorithm to select a personalized subset of these candidate products, based on the personalized scores $r(q, \A_u)$ (Eqn.~\ref{eqn:score}).

\begin{figure}[h!]
  \centering
  \includegraphics[width=0.65\textwidth]{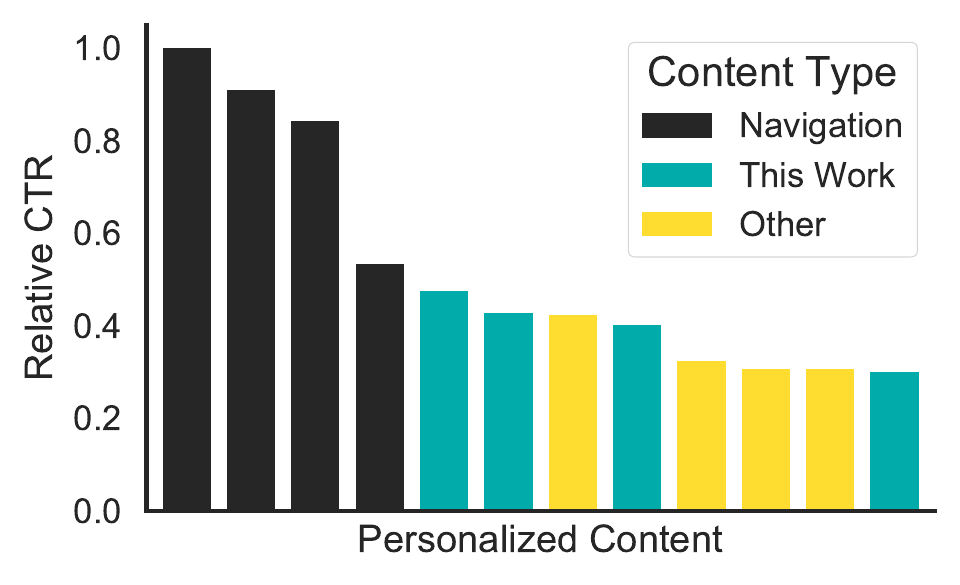}
  \caption{Relative CTR of top-performing personalized content. Each bar corresponds to a single piece of content. Content labeled \emph{This Work} contains product recommendations produced using the model described here, \emph{Other} are product recommendations made using alternate personalization methods, and \emph{Navigation} corresponds to utilitarian content. Depicted CTR is relative to the piece of navigational content with the highest CTR.}
  \label{fig:ctr}
\end{figure}

We evaluated the quality of the resulting pieces of content by considering their click through rate. Figure \ref{fig:ctr} depicts the click through rate (CTR) of top-performing personalized content, relative to the content with the highest CTR. For context, we show this content alongside the highest performing navigational content and existing personalized content on the homepage. Navigational content includes content such as ``Your Orders'' and links to recent searches; unsurprisingly such content has a high CTR as it is used to navigate the site. The other content is produced using various proprietary production personalization algorithms which have been extensively optimized for the task. We see from Figure \ref{fig:ctr} that content produced using the attention-based personalization mechanism compares favorably with content produced using other personalization methods. Indeed, the top two content items in terms of CTR were produced using the attention-based algorithm.

\section{CONCLUSIONS AND FUTURE WORK}

This paper presented an attention based method for leveraing pre-trained item representations for personalized ranking. The proposed method was shown to significantly outperform a number of baselines according to common ranking evaluation metrics using purely behavioral data. Employing this model in an real-world product recommendation setting yielded personalized content that obtained high CTR.

There are many possible variations of the attention based model described here. Exploring pairwise ranking functions, the inclusion of non-behavioral features, extensions to the content ranking setting, reliably adapting item embeddings while learning the rest of the model parameters, personalized candidate selection mechanisms, and better understanding the importance of the temporal nature of item interactions by incorporating recurrent or bidirectional \citep{bidirectional} components into the model are all interesting directions.

Interpretability is also an extremely interesting direction for further research. In contrast to most modern neural machinery, the outputs of attention mechanisms are highly interpretable. For example, in \citet{attention-align-and-translate}, an attention mechanism is used to visualize what source words are most important when generating each target word in a machine translation task. Attention based recommendation systems could provide a mechanism to help users better understand why certain items are being recommended, and help internal merchandisers focus their efforts and create more compelling content.

% \section{Conclusion}

% This paper presenteded \abcf{}, an attention based method for leveraging pre-trained item embeddings for the problem of personalized ranking. The proposed method wass shown to significantly outperform a number of baselines according to common ranking evaluation metrics. Employing \abcf{} in an real-world product recommendation setting yielded personalized content that obtained high CTR. In addition, \abcf{} can easily be extended to incorporate additional features, allowing it to be adapted for domain specific use cases. %We believe this, combined with the availability and ease of using ASIN embeddings, makes our method of general interest to numerous teams within Amazon working on personalization and related problems.

\bibliographystyle{plainnat}
\bibliography{bibliography}

\end{document}